\documentclass[preprint,10pt,pra,showpacs,onecolumn]{revtex4}%
\usepackage{amsfonts}
\usepackage{amsmath}
\usepackage{amssymb}
\usepackage{graphicx}%
\begin{document}
\title{Efficient generation of entangled photons by cavity QED}
\author{X.L. Zhang$^{1,2}$}
\author{M. Feng$^{1}$}
\email{mangfeng1968@yahoo.com}
\author{K.L. Gao$^{1}$}
\affiliation{$^{1}$State Key Laboratory of Magnetic Resonance and Atomic and Molecular
Physics, Wuhan Institute of Physics and Mathematics, Chinese Academy of
Sciences, Wuhan 430071, China}
\affiliation{$^{2}$Graduate School of the Chinese Academy of Science, Bejing 100049, China}


\begin{abstract}
A potential scheme is proposed to generate complete sets of entangled photons
in the context of cavity quantum electrodynamics (QED). The scheme includes
twice interactions of atoms with cavities, in which the first interaction is
made in two-mode optical cavities and the second one\ exists in a microwave
cavity. In the optical cavities the atoms are resonant with the cavity modes,
while the detuned interaction of the atoms with a single-mode of the microwave
cavity is driven by a classical field. We show that our scheme is carried out
with higher efficiency than previeous schemes, and is close to the reach of
current technique.

\end{abstract}
\maketitle

Entanglement is not only an essential resource for quantum information
processing, such as quantum key distribution \cite{1}, quantum dense coding
\cite{2} and quantum teleportation \cite{3}, but also a key ingredient for the
test of quantum nonlocality \cite{4,5,6}. Many current efforts are paid on the
controlled generation and detection of entangled states. Among all physical
realizations of qubits, photons constitute one of the most promising systems
for implementation of quantum information science, because the polarized
states of the photons are individually controllable and their quantum
coherence can be preserved over kilometers in an optical fiber. Most
experimental realizations to generate entangled photons come almost
exclusively from parametric down-conversion in nonlinear crystals \cite{7}.
Besides, one can also entangle photons by making use of atomic cascade decay
\cite{8} or excitonic emission in semiconductor quantum dots \cite{9,10,11,12}.

Cavity QED is an excellent technique to achieve few-qubit entanglement and
quantum computing, for example, Bell state preparation \cite{13} and two-qubit
quantum gates \cite{14}. Recently there have been some proposals to generate
entangled photons in cavity QED \cite{15,16}. In this paper, we propose an
alternative scheme for deterministically generating entangled photons in
cavity QED by two steps for projecting the entanglement from atomic states to
the photons emitted from these atoms. In the first step we generate photons by
sending atoms through an array of optical cavities with each of two
orthogonally polarized modes. In the second step the atoms are entangled by a
detuned interaction with a single-mode microwave cavity. The favorable
features of our scheme include: (1) It is very straightforward in
implementation because we carry out the scheme by only sending atoms through
the cavities. The requirement for the implementation is very close to the
reach of current cavity QED techniques. (2) The complete set of the entangled
two- or more-photon states can be generated deterministically by our scheme,
and the implementation time remains constant with the size of the entangled
photon states. (3) Our scheme is more efficient than previous proposals with
cavities \cite{15,16}, and the generated photons may be collected much more
efficiently, due to cavities, than previous proposals by spontaneous emission
\cite{10,11,12}.

We first consider the simplest case, i.e., creating the EPR photon pairs
$\left\vert \phi^{\pm}\right\rangle =\frac{1}{\sqrt{2}}\left(  \left\vert
\sigma^{+}\sigma^{+}\right\rangle \pm\left\vert \sigma^{-}\sigma
^{-}\right\rangle \right)  $\ and $\left\vert \psi^{\pm}\right\rangle
=\frac{1}{\sqrt{2}}\left(  \left\vert \sigma^{+}\sigma^{-}\right\rangle
\pm\left\vert \sigma^{-}\sigma^{+}\right\rangle \right)  $.\ As shown in Fig.
1(a), we consider two atoms going through two optical cavities (i.e., cavities
1 and 2), respectively and resonantly interacting with the cavity modes. Then
they simultaneously go through a single-mode microwave cavity (i.e., the
cavity 3) by detuned interaction with the cavity mode, driven by a classical
field. Each of the atoms consists of two ground levels $\left\vert
g\right\rangle $ and $\left\vert e\right\rangle $, and an excited level
$\left\vert r\right\rangle $ (See Fig. 1(b)).

Let us first consider the first step of the scheme regarding the optical
cavities. We assume that the cavities are initially empty and the atoms are in
state $\left\vert r\right\rangle $, i.e., $\left\vert \psi(0)\right\rangle
_{j}=\left\vert r,0_{L},0_{R}\right\rangle _{j}$ with $\left\vert
\cdots\right\rangle _{j}$ denoting the atomic state, the left and right modes
of the optical cavity j, respectively. The initial preparation of the atomic
states in exited states $\left\vert r\right\rangle _{j}$ could be made by
pumping from the corresponding ground states by some resonant $\pi
$-polarization lasers before the scheme gets started. Under the rotating-wave
approximation, we have following Hamiltonian in units of $\hbar=1$,%
\begin{align}
H_{j}  &  =\frac{1}{2}\left[  \left(  h_{jR}a_{jR}^{+}\left\vert
g\right\rangle _{jj}\left\langle r\right\vert +h_{jL}a_{jL}^{+}\left\vert
e\right\rangle _{jj}\left\langle r\right\vert \right)  +h.c.\right]
\nonumber\\
&  -i\frac{\gamma}{2}\left\vert r\right\rangle _{jj}\left\langle r\right\vert
-i\kappa_{j}(a_{jL}^{+}a_{jL}+a_{jR}^{+}a_{jR}),
\end{align}
where $h_{jR}$ and $h_{jL}$ are, respectively, the coupling strengths of the
jth atom to the right-circularly ($\sigma^{+}$) and left-circularly
($\sigma^{-}$) polarized modes of the cavity field. $a_{jL}$ and $a_{jR}$
($a_{jL}^{+}$ and $a_{jR}^{+}$) are the annihilation (creation) operators of
the left-circularly and right-circularly polarized modes of the jth cavity,
respectively. $\gamma$ is regarding the spontaneous emission of the excited
state $\left\vert r\right\rangle $ and $\kappa_{j}$ accounts for the decay
rate of the jth cavity. For simplicity, we assume $\kappa_{jR}=\kappa
_{jL}=\kappa_{j}$ and $\gamma_{1}=\gamma_{2}=\gamma$. Before any photon leaks
out of each cavity, we may solve Eq. (1) following the solutions in
\cite{16,17}. After an interaction time $\tau_{j}$, the system evolves to the
entangled state,%
\begin{equation}
\left\vert \psi(t)\right\rangle _{j}=\frac{\exp(-\frac{\kappa_{j}+\frac
{\gamma}{2}}{2}\tau_{j})}{2\Omega_{j}}(a\left\vert r,0_{L},0_{R}\right\rangle
_{j}+b\left\vert g,0_{L},1_{R}\right\rangle _{j}+c\left\vert e,1_{L}%
,0_{R}\right\rangle _{j}),
\end{equation}
with%
\begin{align*}
a  &  =[(\kappa_{j}-\frac{\gamma}{2})\sin(\Omega_{j}\tau_{j})+2\Omega_{j}%
\cos(\Omega_{j}\tau_{j})],\\
b  &  =-ih_{jR}\sin(\Omega_{j}\tau_{j}),\\
c  &  =-ih_{jL}\sin(\Omega_{j}\tau_{j}),\\
\Omega_{j}  &  =\frac{1}{2}\sqrt{2\gamma\kappa_{j}+h_{jR}^{2}+h_{jL}%
^{2}-\left(  \kappa_{j}+\frac{\gamma}{2}\right)  ^{2}}.
\end{align*}

We may carefully control the atoms' velocities to make the interaction time
$\tau_{j}$ satisfy $\tan(\Omega_{j}\tau_{j})=2\Omega_{j}/(\frac{\gamma}%
{2}-\kappa_{j})$, which yields Eq. (2) reducing to%
\begin{equation}
\left\vert \psi\right\rangle _{j}=\frac{1}{\sqrt{h_{jR}^{2}+h_{jL}^{2}}%
}\left(  h_{jR}\left\vert g,0_{L},1_{R}\right\rangle _{j}+h_{jL}\left\vert
e,1_{L},0_{R}\right\rangle _{j}\right)  ,
\end{equation}
with the success probability%
\begin{equation}
P_{j}=\exp\left[  -\left(  \kappa_{j}+\frac{\gamma}{2}\right)  \tau
_{j}\right]  \sin^{2}(\Omega_{j}\tau_{j})\left(  \sqrt{h_{jR}^{2}+h_{jL}^{2}%
}/2\Omega_{j}\right)  ^{2}.
\end{equation}
When $t>max\{1/(\kappa_{1}+\frac{\gamma}{2}),$ $1/(\kappa_{2}+\frac{\gamma}%
{2})\},$ photons will leak out of the cavities, and the total system evolves
into%
\begin{equation}
\left\vert \Phi\right\rangle =%
{\textstyle\prod_{j=1}^{2}}
\frac{1}{\sqrt{h_{jR}^{2}+h_{jL}^{2}}}\left(  h_{jR}\left\vert g\right\rangle
_{j}\left\vert \sigma^{+}\right\rangle +h_{jL}\left\vert e\right\rangle
_{j}\left\vert \sigma^{-}\right\rangle \right)  .
\end{equation}

To entangle the two emitted photons based on Eq. (5), we must entangle the two
atoms, which is the task of the second step. Consider that the two atoms get
out of the optical cavities 1 and 2, respectively, and then simultaneously get
in the cavity 3, which is a single-mode microwave cavity. We suppose that the
two atoms are detuned from the mode of the cavity 3, driven by a classical
field. The Hamiltonian is \cite{18}
\begin{equation}
H_{2}=\omega_{0}\sum_{j=1,2}S_{z,j}+\omega_{c}a^{+}a\text{ }+G\sum
_{j=1,2}\left(  e^{-i\omega t}S_{j}^{+}+e^{i\omega t}S_{j}^{-}\right)
+g\sum_{j=1,2}\left(  S_{j}^{+}a+S_{j}^{-}a^{+}\right)  ,
\end{equation}
where $S_{z,j}=\left(  \left\vert e\right\rangle _{jj}\left\langle
e\right\vert -\left\vert g\right\rangle _{jj}\left\langle g\right\vert
\right)  /2$, $S_{j}^{+}=\left\vert e\right\rangle _{jj}\left\langle
g\right\vert $, and $S_{j}^{-}=\left\vert g\right\rangle _{jj}\left\langle
e\right\vert $, with $\left\vert e\right\rangle _{j}$ and $\left\vert
g\right\rangle _{j}$\ (j=1 and 2) being states in Fig. 1(b) of the jth atom.
$a^{+}$ and $a$ are the creation and annihilation operators for the microwave
cavity mode, respectively. $\omega_{0}$, $\omega_{c}$ and $\omega$ are,
respectively, frequencies\ with respect to resonant transition between levels
$\left\vert e\right\rangle \ $and $\left\vert g\right\rangle $, to the
microwave cavity and to the classical driving field. $g$ and $G$ are the
coupling constants of each atom to the cavity mode and to the driving field,
respectively. We assume that $\omega_{0}=\omega$, and the rotating-wave
approximation yields an effective Hamiltonian in the rotating framework with
respect to the interaction-free part of Eq. (6),%
\begin{equation}
H_{2}^{^{\prime}}=G\sum_{j=1,2}\left(  S_{j}^{+}+S_{j}^{-}\right)
+g\sum_{j=1,2}\left(  e^{i\delta t}S_{j}^{+}a+e^{-i\delta t}S_{j}^{-}%
a^{+}\right)  ,
\end{equation}
where $\delta=\omega_{0}-\omega_{c}$. By defining the dressed states
$\left\vert \pm\right\rangle _{j}=\frac{1}{\sqrt{2}}\left(  \left\vert
g\right\rangle _{j}\pm\left\vert e\right\rangle _{j}\right)  $, we make a
further rotating transformation with respect to the terms regarding $G$ in Eq.
(7), and obtain%
\begin{equation}
H_{2}^{i}=\frac{g}{2}\sum_{j=1,2}(\left\vert +\right\rangle _{jj}\left\langle
+\right\vert -\left\vert -\right\rangle _{jj}\left\langle -\right\vert
+e^{i2Gt}\sigma_{j}^{+}-e^{-i2Gt}\sigma_{j}^{-})e^{i\delta t}a+H.c.,
\end{equation}
where $\sigma_{j}^{+}=\left\vert +\right\rangle _{jj}\left\langle -\right\vert
,$ and $\sigma_{j}^{-}=\left\vert -\right\rangle _{jj}\left\langle
+\right\vert .$ In the strong driving regime $G\gg\{\delta,g\}$, we may
neglect the fast oscillating terms regarding $e^{\pm i2Gt}$ and get%
\[
H_{2}^{I}=\frac{g}{2}\sum_{j=1,2}(S_{j}^{+}+S_{j}^{-})(e^{i\delta
t}a+e^{-i\delta t}a^{+}),
\]
In the case of $\delta\gg g/2$ there is no energy exchange between the atomic
system and the cavity. So the possible energy-conserving transitions are
between the states $\left\vert e\right\rangle _{1}\left\vert g\right\rangle
_{2}\left\vert n\right\rangle $ and $\left\vert g\right\rangle _{1}\left\vert
e\right\rangle _{2}\left\vert n\right\rangle ,$ and between $\left\vert
e\right\rangle _{1}\left\vert e\right\rangle _{2}\left\vert n\right\rangle $
and $\left\vert g\right\rangle _{1}\left\vert g\right\rangle _{2}\left\vert
n\right\rangle $, mediated by virtually excited cavity states $\left\vert
n\pm1\right\rangle $. Since the transition paths in both $\left\vert
g\right\rangle _{1}\left\vert e\right\rangle _{2}\left\vert n\right\rangle $
$\leftrightarrow$ $\left\vert e\right\rangle _{1}\left\vert g\right\rangle
_{2}\left\vert n\right\rangle $ and $\left\vert e\right\rangle _{1}\left\vert
e\right\rangle _{2}\left\vert n\right\rangle $ $\leftrightarrow$ $\left\vert
g\right\rangle _{1}\left\vert g\right\rangle _{2}\left\vert n\right\rangle $
interfere destructively, the effective coupling constant $\lambda$ is
independent of the photon number in the cavity \cite{19},%
\begin{align*}
\lambda &  =2(\frac{_{1}\left\langle g\right\vert _{2}\left\langle
e\right\vert \left\langle n\right\vert H_{2}^{I}\left\vert g\right\rangle
_{1}\left\vert g\right\rangle _{2}\left\vert n+1\right\rangle _{1}\left\langle
g\right\vert _{2}\left\langle g\right\vert \left\langle n+1\right\vert
H_{2}^{I}\left\vert e\right\rangle _{1}\left\vert g\right\rangle
_{2}\left\vert n\right\rangle }{\delta}\\
&  +\frac{_{1}\left\langle g\right\vert _{2}\left\langle e\right\vert
\left\langle n\right\vert H_{2}^{I}\left\vert e\right\rangle _{1}\left\vert
e\right\rangle _{2}\left\vert n-1\right\rangle _{1}\left\langle e\right\vert
_{2}\left\langle e\right\vert \left\langle n-1\right\vert H_{2}^{I}\left\vert
e\right\rangle _{1}\left\vert g\right\rangle _{2}\left\vert n\right\rangle
}{-\delta})\\
&  =g^{2}/2\delta.
\end{align*}
The same coupling constant $\lambda$ can be obtained from the transitions
between $\left\vert e\right\rangle _{1}\left\vert e\right\rangle
_{2}\left\vert n\right\rangle $ and $\left\vert g\right\rangle _{1}\left\vert
g\right\rangle _{2}\left\vert n\right\rangle $, intermediated by virtually
excited cavity states $\left\vert n\pm1\right\rangle $. The Stark shift for
the state $\left\vert e\right\rangle _{j}$ is
\begin{align*}
\lambda^{^{\prime}}  &  =\frac{_{j}\left\langle e\right\vert \left\langle
n\right\vert H_{2}^{I}\left\vert g\right\rangle _{j}\left\vert
n+1\right\rangle _{j}\left\langle g\right\vert \left\langle n+1\right\vert
H_{2}^{I}\left\vert e\right\rangle _{j}\left\vert n\right\rangle }{\delta}\\
&  +\frac{_{j}\left\langle e\right\vert \left\langle n\right\vert H_{2}%
^{I}\left\vert g\right\rangle _{j}\left\vert n-1\right\rangle _{j}\left\langle
g\right\vert \left\langle n-1\right\vert H_{2}^{I}\left\vert e\right\rangle
_{j}\left\vert n\right\rangle }{-\delta}\\
&  =g^{2}/4\delta,
\end{align*}
where j=1 and 2. There same values for the case of $\left\vert g\right\rangle
_{j}$ and then $H_{2}^{I}$ turns to%
\begin{equation}
H_{2}^{eff}=\lambda^{^{\prime}}\sum_{j=1,2}\left(  \left\vert e\right\rangle
_{jj}\left\langle e\right\vert +\left\vert g\right\rangle _{jj}\left\langle
g\right\vert \right)  +\lambda\text{ }\left(  S_{1}^{+}S_{2}^{+}+S_{1}%
^{+}S_{2}^{-}+H.c.\right)  ,
\end{equation}
where the effective coupling constant $\lambda$ and $\lambda^{^{\prime}}$ are
both independent of the photon number of the cavity field. If the two atoms
are in the state in Eq. (5) before getting in the microwave cavity, a
straightforward algebra leads to the evolved state of the total system
\cite{20},%
\begin{align}
&  \widetilde{N}[\left\vert gg\right\rangle (h_{1R}h_{2R}\left\vert \sigma
^{+}\sigma^{+}\right\rangle -ih_{1L}h_{2L}\left\vert \sigma^{-}\sigma
^{-}\right\rangle )\nonumber\\
&  -i\left\vert ee\right\rangle (h_{1R}h_{2R}\left\vert \sigma^{+}\sigma
^{+}\right\rangle +ih_{1L}h_{2L}\left\vert \sigma^{-}\sigma^{-}\right\rangle
)\nonumber\\
&  +\left\vert ge\right\rangle (h_{1R}h_{2L}\left\vert \sigma^{+}\sigma
^{-}\right\rangle -ih_{1L}h_{2R}\left\vert \sigma^{-}\sigma^{+}\right\rangle
)\\
&  -i\left\vert eg\right\rangle (h_{1R}h_{2L}\left\vert \sigma^{+}\sigma
^{-}\right\rangle +ih_{1L}h_{2R}\left\vert \sigma^{-}\sigma^{+}\right\rangle
)],\nonumber
\end{align}
where $\widetilde{N}=\frac{1}{\sqrt{2}\sqrt{h_{1R}^{2}+h_{1L}^{2}}\sqrt
{h_{2R}^{2}+h_{2L}^{2}}}$ is the normalization constant, and we have chosen
$\lambda t=\pi/4$, and $Gt=n\pi,$ with n being an integer. For simplicity, we
only consider the case of $h_{jR}=h_{jL}=h_{j}$. So we have
\begin{align}
&  (1/2\sqrt{2})[\left\vert gg\right\rangle (\left\vert \sigma^{+}\sigma
^{+}\right\rangle -i\left\vert \sigma^{-}\sigma^{-}\right\rangle )-i\left\vert
ee\right\rangle (\left\vert \sigma^{+}\sigma^{+}\right\rangle +i\left\vert
\sigma^{-}\sigma^{-}\right\rangle )\nonumber\\
&  -i\left\vert eg\right\rangle (\left\vert \sigma^{+}\sigma^{-}\right\rangle
+i\left\vert \sigma^{-}\sigma^{+}\right\rangle )+\left\vert ge\right\rangle
(\left\vert \sigma^{+}\sigma^{-}\right\rangle -i\left\vert \sigma^{-}%
\sigma^{+}\right\rangle )].
\end{align}
Eq. (11) presents a deterministic generation of entangled photon pairs with
our scheme, while to obtain a certain entangled photon pairs, we only have the
success rate 1/4. Moreover, the classical driving field can be removed from
the second step in above implementation, which actually corresponds to the
experimental implementation in \cite{13}. But in that case, only $\left\vert
\psi^{\pm}\right\rangle $, instead of a complete set of Bell states in Eq.
(11), could be obtained.

Our scheme is suitable for not only repeatedly producing entangled photon
pairs, but also generating entangled multiphotons. For the latter case, after,
for example, N atoms are sent through N two-mode optical cavities, i.e.,
cavities 1, 2, $\cdots$, N\ in Fig. 1(c), respectively,\ a careful control of
the interaction time $\tau_{j}$ to satisfy $\tan(\Omega_{j}\tau_{j}%
)=2\Omega_{j}/(\frac{\gamma}{2}-\kappa_{j})$ would lead to a state,%
\begin{equation}
\left\vert \Phi\right\rangle =%
{\textstyle\prod_{j=1}^{N}}
\frac{1}{\sqrt{2}}\left(  \left\vert g\right\rangle _{j}\left\vert \sigma
^{+}\right\rangle +\left\vert e\right\rangle _{j}\left\vert \sigma
^{-}\right\rangle \right)  ,
\end{equation}
where we have assumed $h_{jR}=h_{jL}=h_{j}$ (j=1, 2, $\cdots$, N) for
simplicity. A convenient treatment for an ensemble of spin-1/2 atoms is to use
the collective spin operators. By transforming the atomic states to
eigenstates of a collective operator $S_{x}$ with%

\[
S_{x}=\frac{1}{2}\sum_{j=1}^{N}(S_{j}^{+}+S_{j}^{-}),
\]
we may get the following effective Hamiltonian in the interaction picture
related to Eq. (6),%
\begin{equation}
H=(i\frac{\partial}{\partial t}U)U^{+}+UH_{N}^{eff}U^{+}=2GS_{x}+2\lambda
S_{x}^{2},
\end{equation}
where $U=e^{-i2GS_{x}t},$ and $H_{N}^{eff}$ is of similar form to $H_{2}%
^{eff}$ but for N atoms. We have noticed a recent work \cite{19} for
entangling many trapped ions by using angular momentum representation with
rotating basis states $\left\vert N/2,M\right\rangle _{x}$, M = -N/2, ....,
N/2, with respect to the atomic states. Since $\left\vert N/2,M\right\rangle
_{x}$are the eigenstates of $S_{x},$\ by using Eq. (13), we could obtain the
analytical form of the entangled ionic states by direct algebra. In our case,
however, due to the degenerate states, the basis states above are not
sufficient for a complete subspace. So we have to introduce an additional
degree of freedom K to lift the degeneracy, and\ thereby the basis states are
$\left\vert N/2,M,K\right\rangle _{x}$ with M = -N/2, ...., N/2 and K=0,1,
....N. The newly introduced K accounts for the number of the minus signs
$^{\prime}$-$^{\prime}$ contributed from the excited atomic levels. So for a
state with $n_{e}$ components regarding the excited atomic level,\ we have%
\begin{equation}
|g...g\underset{n_{e}}{\underbrace{e...e}}\rangle=%
{\textstyle\prod_{M,K}}
\text{ }C_{MK}\text{ }(-1)^{N/2-M}\text{ }\left\vert N/2,M,K\right\rangle
_{x},
\end{equation}
where M = -N/2, ...., N/2, K=0,1,...,$n_{e},$ $C_{MK}=(-1)^{K}C_{M}$ and
$C_{M}$ is given in \cite{19}. A complete discussion about the properties of
this full set of basis states could be found in \cite{21}. In the present
paper, however, our interest is only in the generation of entangled states
based on these states. Considering current experimental feasibility and also
for clarity, we will below demonstrate a generation of three entangled photons
in detail by a formally simpler solution than in \cite{21}. For the
Hamiltonian Eq. (13) with N=3, we transform the atomic states into the
eigenstates of $S_{x}$, and obtain,%
\begin{equation}
\left(
\begin{array}
[c]{c}%
\left\vert ggg\right\rangle \\
\left\vert gge\right\rangle \\
\left\vert geg\right\rangle \\
\left\vert gee\right\rangle \\
\left\vert egg\right\rangle \\
\left\vert ege\right\rangle \\
\left\vert eeg\right\rangle \\
\left\vert eee\right\rangle
\end{array}
\right)  =\frac{1}{8}\left(
\begin{array}
[c]{cccccccc}%
1 & 1 & 1 & 1 & 1 & 1 & 1 & 1\\
1 & -1 & 1 & -1 & 1 & -1 & 1 & -1\\
1 & 1 & -1 & -1 & 1 & 1 & -1 & -1\\
1 & -1 & -1 & 1 & 1 & -1 & -1 & 1\\
1 & 1 & 1 & 1 & -1 & -1 & -1 & -1\\
1 & -1 & 1 & -1 & -1 & 1 & -1 & 1\\
1 & 1 & -1 & -1 & -1 & -1 & 1 & 1\\
1 & -1 & -1 & 1 & -1 & 1 & 1 & -1
\end{array}
\right)  \left(
\begin{array}
[c]{c}%
\left\vert +++\right\rangle \\
\left\vert ++-\right\rangle \\
\left\vert +-+\right\rangle \\
\left\vert +--\right\rangle \\
\left\vert -++\right\rangle \\
\left\vert -+-\right\rangle \\
\left\vert --+\right\rangle \\
\left\vert ---\right\rangle
\end{array}
\right)  .
\end{equation}
By choosing $Gt=\left(  2n+3/4\right)  \pi$\ and $\lambda t=\pi/4$, we have
following evolution of the three-atom states.%
\begin{align}
\left\vert ggg\right\rangle  &  \Longrightarrow\frac{1}{\sqrt{2}}(\left\vert
ggg\right\rangle +i\left\vert eee\right\rangle ),\text{ }\left\vert
gge\right\rangle \Longrightarrow\frac{1}{\sqrt{2}}(\left\vert gge\right\rangle
+i\left\vert eeg\right\rangle ),\nonumber\\
\left\vert geg\right\rangle  &  \Longrightarrow\frac{1}{\sqrt{2}}(\left\vert
geg\right\rangle +i\left\vert ege\right\rangle ),\text{ }\left\vert
gee\right\rangle \Longrightarrow\frac{1}{\sqrt{2}}(\left\vert gee\right\rangle
+i\left\vert egg\right\rangle ),\nonumber\\
\left\vert egg\right\rangle  &  \Longrightarrow\frac{1}{\sqrt{2}}(\left\vert
egg\right\rangle +i\left\vert gee\right\rangle ),\text{ }\left\vert
ege\right\rangle \Longrightarrow\frac{1}{\sqrt{2}}(\left\vert ege\right\rangle
+i\left\vert geg\right\rangle ),\nonumber\\
\left\vert eeg\right\rangle  &  \Longrightarrow\frac{1}{\sqrt{2}}(\left\vert
eeg\right\rangle +i\left\vert gge\right\rangle ),\text{ }\left\vert
eee\right\rangle \Longrightarrow\frac{1}{\sqrt{2}}(\left\vert eee\right\rangle
+i\left\vert ggg\right\rangle ),
\end{align}
where we have discarded their common global phase. Therefore, if the atoms are
in a state in Eq. (12) with N=3, after they go through the microwave cavity
simultaneously, a certain detection on the atomic internal states would yield
a certain state from the complete set of three-photon entangled states.

Following above algebra, we may have a similar state evolution to Eq. (16),
after lengthy but straightforward deduction, for N atoms prepared in Eq. (12)
and sent through the microwave cavity simultaneously, under the conditions
$Gt=\left(  2n+3/4\right)  \pi$\ and $\lambda t=\pi/4$ when N is odd, or under
the conditions $Gt=n\pi$\ and $\lambda t=\pi/4$ when N is even.

To get a highly efficient generation of entangled photons, we require
$\frac{\gamma}{2}<\kappa_{j}<h_{j}$. In optical cavities, the single-photon
coupling strength $h=2\pi\times34$ MHz, the atomic decay rate $\gamma=$
$2\pi\times2.6$ MHz and the cavity decay rate $\kappa=2\pi\times4.1$ MHz have
been reported \cite{22,23}. So in the case of two atoms, by setting
$\kappa_{1}=\kappa_{2}=\kappa$, and $h_{1}=h_{2}=h,$ we get the success
probability to be $P=P_{1}\times P_{2}=48.1\%$ at $\tau_{j}=10.8$ ns. The
efficiency can be higher with smaller parameters $\gamma$ and $\kappa,$ as
shown in Fig. 2. In the second step, we assume the coupling of the atoms with
the microwave cavity to be $g\simeq2\pi\times50kHz$, and the photon storage
time in a microwave cavity to be $T_{c}=1$ ms (corresponding to Q=$3\times
10^{8}$) \cite{24}. So with the choice $\delta=5g,$ the interaction time of
the atom with the cavity field is of the order of tenth of $micro\sec$.

Due to the large detuning employed in the second step, the success rate of the
photon generation under the detrimental influence of the cavity decay in our
scheme is only related to the optical cavities. In this sense, we may make a
comparison of our operations in the optical cavity with those in \cite{15}.
Because of the near resonance and also because the generation of the second
photon is based on that of the first photon, a relatively big cavity decay
rate is required in \cite{15}. In contrast, our scheme goes better with
smaller cavity decay. So it is understandable that our scheme is of much
higher success rate than that (i.e., approximate 0.06) in \cite{15} (See Fig.
2). More importantly, our scheme can be directly extended to the preparation
of many-photon entanglement, while this is impossible in \cite{15}. The recent
proposal based on a magnetic field gradient also enables a generation of
multiphoton entangled states by mapping entanglement from atoms to photons.
But it is much slower than our present scheme. It generates an entangled
photon pair by a time of the order of milisec, and the implementation time
would be much longer for producing entangled states of more photons. In
contrast, no matter how many photons would be entangled by our scheme, the
implementation time remains constant.

To the best of our knowledge two atoms interacting coherently in a microwave
cavity have been achieved experimentally \cite{13}, while we have not yet
found any experimental report for more than two atoms controllably interacting
in a microwave cavity. A big challenge for an experimental realization of our
scheme is to sent the atoms through the microwave cavity simultaneously, and
any deviation from the simultaneousness would lead to infidelity. We have
assessed the infidelity due to operational imperfection in Fig. 3 for the
cases of two and three atoms. For simplicity, we suppose that the atoms are
moving with the same speed, but enter the microwave cavity sequentially with
the time difference $\delta t$ between neighboring atoms. For the two-atom
case, the interaction time of the two atoms with the detuned cavity mode is
thereby reduced to $(t_{0}-2\delta t)$ with $t_{0}$ the desired time in the
ideal case. In addition to the single atom resonantly interacting with the
classical driving field during the time interval $\delta t$, as well as free
evolution of the atoms, we can obtain the infidelity, due to the operational
imperfection, to be $\sin^{2}(G\delta t)+\cos^{2}(G\delta t)\{1$ $-$
$\sin[2\lambda(t-2\delta t)]\}/2.$ Similar consideration on the three-atom
case yields a more complicated analytical result for the infidelity which
omitted here. \ The numerical calculation in Fig. 3 could tell us that the
infidelity would be increasing with the atom number and $\delta t,$ while our
scheme works well if $\delta t<0.01t_{0}$ in cases of N=2 and 3. On the other
hand, in the case of the atoms with different speeds, the infidelity is also
obtainable, similar to the result to the above discussion for non-simultaneous
movement of the atoms, if the two atoms get in the microwave cavity at the
same time. As one atom will go out of the microwave cavity before the other
one, the time deviation from the desired time will yield a single atom
resonantly interacting with the classical driving field. But if the atoms
going through the respective optical cavities with different times from the
desired ones, i.e., not meeting the condition $\tan(\Omega_{j}\tau
_{j})=2\Omega_{j}/(\frac{\gamma}{2}-\kappa_{j})$, then additional infidelity
will be yielded. So, in principle, the situation for atoms with different
speeds is worse than that for atoms with the same speed but without
simultaneous movement in the microcavity.

It is evident that our scheme is still challenging experimentally. First of
all, our requirement for optical cavities with two orthogonal modes of
different frequencies has not yet achieved experimentally so far. But we have
noticed significant advances in recent experiments \cite{22,23} with optical
cavities including strong coupling of atoms and identification of individual
atoms, which implies that a single atom going through an optical cavity is
controllable. Secondly, for atoms sent through a microwave cavity
simultaneously, we have noticed that, even in the two-atom case, the achieved
experiment \cite{13} was done by using van der Waals collision of the atoms in
the central area of the microwave cavity, instead of sending the atoms through
the microwave cavity strictly simultaneously. While this experimental
discrepancy from the theoretical design did not hamper\ further proposals for
quantum information processing with cavity QED based on simultaneously sending
many atoms through a cavity \cite{14,25}. Like those proposals, we propose
this scheme also based on the expectation that the above mentioned
difficulties in experiments would be overcome in the future with more advances
in cavity QED techniques.

As a final remark, we emphasize that the photons entangled in polarization but
with different energies should be as useful as those entangled identical
photons in quantum information processing, if our implementation is only on
the polarized degrees of freedom. A previous scheme based on biexcitons in
semiconductor quantum dots also produces entanglement of two photons with
different energies \cite{9}.

In conclusion, we have proposed a potential scheme for creating complete sets
of entangled two- or more-photon states in the context of cavity QED, which is
more efficient than recent proposals \cite{15,16}. In contrast to previous
proposals for generating entangled photons by spontaneous emission, our scheme
carried out by means of cavities would\ be of much higher rate for photon
collection. This is of significant importance in view of the application of
entangled photon source. Moreover, our scheme enables a deterministic
generation of entangled multiphoton states, which is also more efficient than
by stochastic method with parametric down-conversion. More importantly, our
scheme is close to the reach of current techniques of cavity QED.

This work is supported by National Natural Science Foundation of China under
Grants No. 10474118 and No. 10274093, by Hubei provincial funding for
distinguished youth, and by the National Fundamental Research Program of China
under Grant No. 2005CB724502.

\begin{figure}
[ptb]
\begin{center}
\includegraphics[
height=3.8977in,
width=3.8787in
]%
{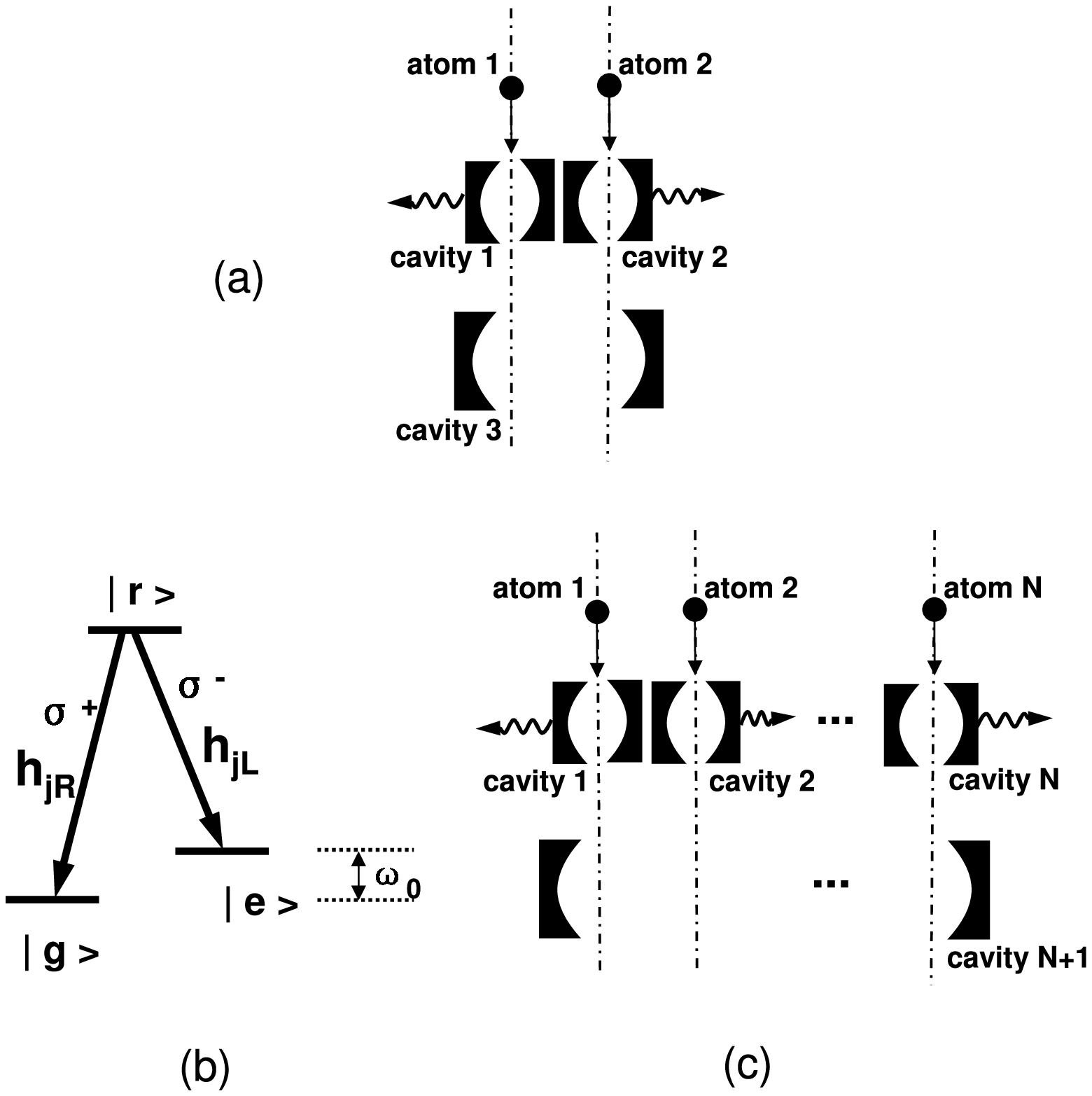}%
\caption{Fig1}%
\end{center}
\end{figure}
\begin{figure}
[ptbptb]
\begin{center}
\includegraphics[
height=7.0534in,
width=4.9035in
]%
{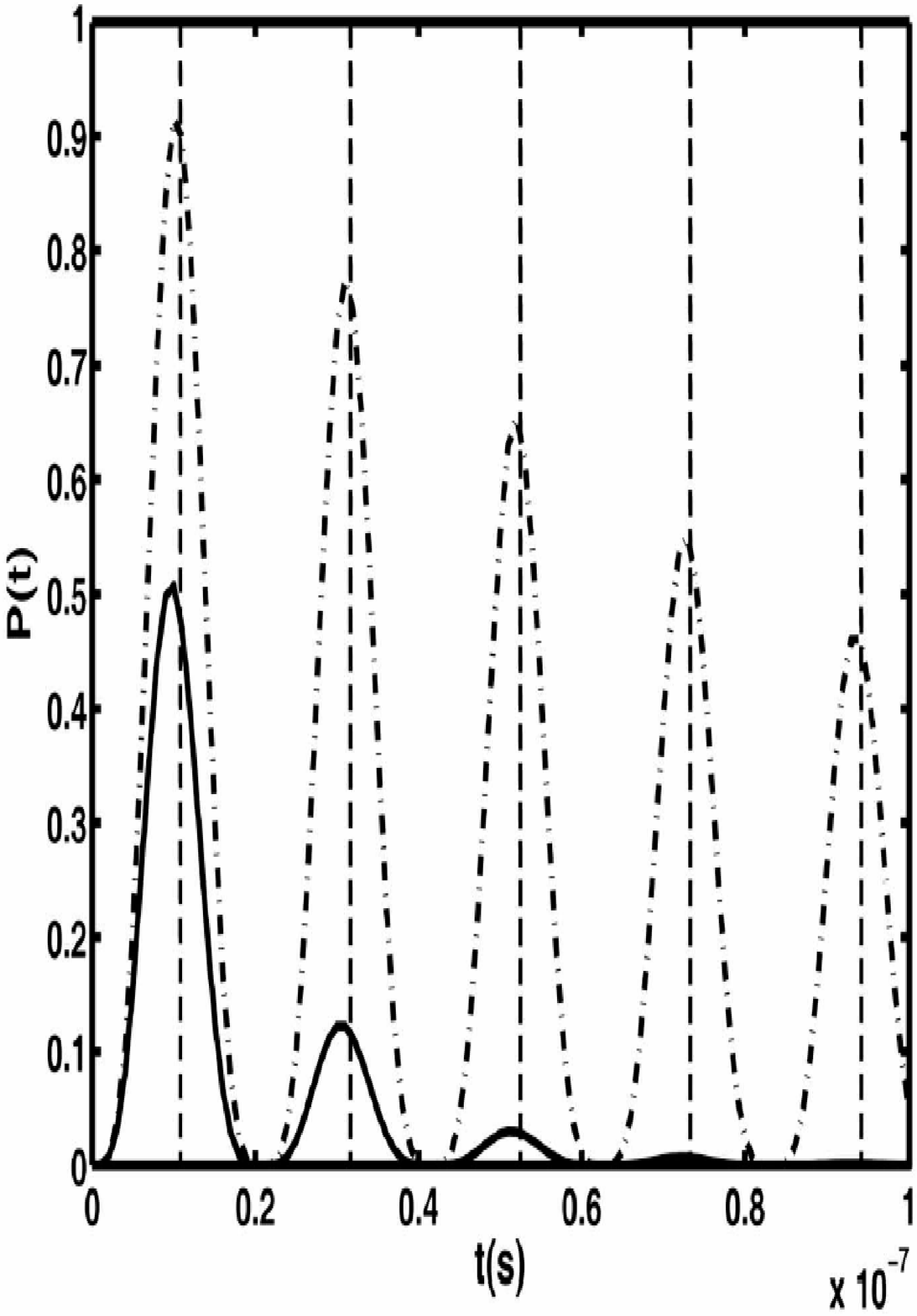}%
\caption{Fig2}%
\end{center}
\end{figure}
\begin{figure}
[ptbptbptb]
\begin{center}
\includegraphics[
height=7.0534in,
width=4.9035in
]%
{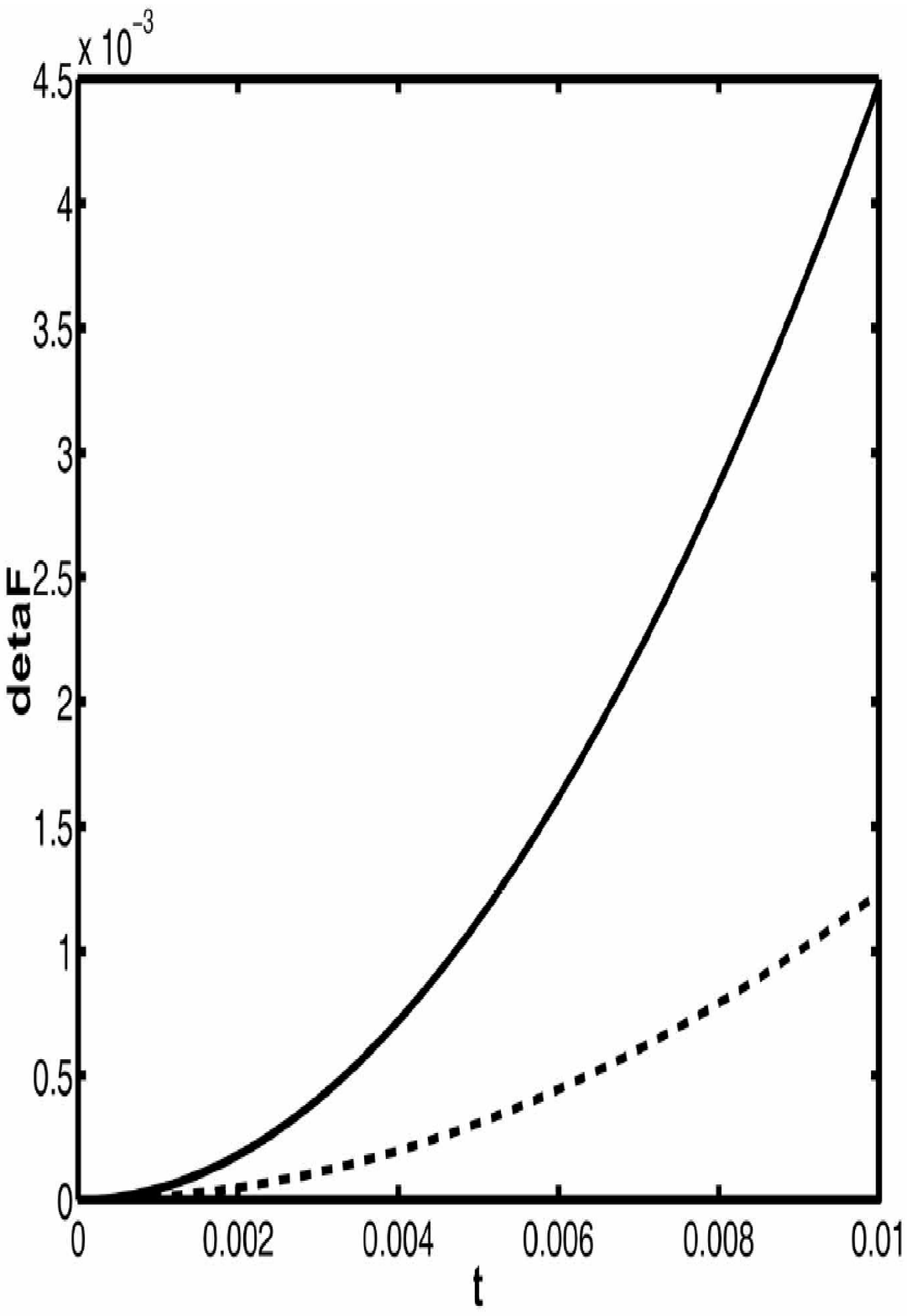}%
\caption{Fig3}%
\end{center}
\end{figure}


\begin{thebibliography}{99}
\bibitem {20}Eq. (10) and the following Eq. (16) are actually written in the
interacting picture related to Eq. (7). If returning to the
Schr\"{o}dinger\ picture, we have different phase factors between $\left\vert
e\right\rangle $ and $\left\vert g\right\rangle $, which have no effect on the
main result due to our detection on the atomic levels.
\end{thebibliography}
\end{document}